\documentclass[12pt]{iopart}

\begin{document}

\title[The stability of exfoliated FeSe nanosheets during in-air device fabrication processes]{The stability of exfoliated FeSe nanosheets during in-air device fabrication processes}

\author{Rui Yang$^2$, Weijun Luo$^1$,  Shun Chi$^2$, Douglas Bonn$^2$ and Guangrui (Maggie) Xia$^{1,\star}$}

\address{$^1$the University of British Columbia, Department of Materials Engineering, Vancouver, B.C., V6T1Z4, Canada}
\address{$^2$the University of British Columbia, Department of Physics and Astronomy, Vancouver, B.C., V6T1Z4, Canada}
%\address{$^3$the University of Illinois at Urbana-Champaign, Department of Electrical and Computer Engineering, Champaign, IL, 61801, USA}

\ead{gxia@mail.ubc.ca}

\vspace{10pt}
\begin{indented}
\item[]February 2018
\end{indented}

\begin{abstract}
We studied the stability and superconductivity of FeSe nanosheets during an in-air device fabrication process. Methods were developed to improve the exfoliation yield and to maintain the superconductivity of FeSe. Raman spectroscopy, atomic force microscopy, optical microscopy and time-of-flight-secondary-ion-mass-spectroscopy measurements show that FeSe nanosheets decayed in air. Precipitation of Se particles and iron oxidation likely occurred during the decay process. Transport measurements revealed that the superconductivity of FeSe disappeared during a conventional electron beam lithography process. Shadow mask evaporation and transfer onto pre-defined electrodes methods were shown to be effective in maintaining the superconductivity after the in-air device fabrication process. These methods developed provide a way of making high quality FeSe nano-devices.
\end{abstract}

\section{Introduction}

Within the last several years, layered transition metal chalcogenides compounds (TMX) have drawn more and more attention. Among these compounds, a spectrum of different behaviours were observed. FeSe is a superconducting member of this family. The reduced dimension and much lower areal carrier density make the exfoliated superconductor FeSe nanosheets ideal materials for the study of tunable superconductivity and quantum phase transition\cite{QPT1, QPT2,QPT2b, QPT3, sit, sit2}. FeSe also belongs to another mystery and intriguing family --- the Fe-based superconductors\cite{fesc0,fesc02,fesc03,fesc04,fesc05,fesc06}. This superconductor family is believed to host the unconventional spin-fluctuation pairing mechanism\cite{fese1,fesc1}. As the simplest member in this family, FeSe is regarded as a key to understand the behaviors of this family\cite{fesc03}. Monolayer FeSe grown on SrTiO$_3$ substrates by Molecular Beam Epitaxy (MBE) even reached a superconductivity transition temperature $T_{c}$ above $100K$\cite{fesc2}. However, MBE-grown FeSe films are commonly affected by the substrates and thus are strained and doped. FeSe films made by top-down approaches are attached to the substrates by weak Van de Waals bonding, and we expect the superconductivity and electronic properties of those films are closer to that of bulk FeSe materials. Could the FeSe thin films made by a top-down approach also reach such a high transition temperature? Can we tune the $T_{c}$ of the FeSe films to observe the quantum phase transition in 2d? The fabrication of FeSe thin films from a top-down approach is the first step towards addressing these questions. So far, there have only been a handful reports on the exfoliated FeSe films \cite{fese1}. Moreover, no studies are available on the stability of FeSe films, which is a common concern for 2d materials in TMX family.

In this work, we first studied the exfoliation of FeSe nanosheets and an effective method to improve the yield of nanosheets was demonstrated. The air instability of FeSe nanosheets was observed and further investigated by the Raman spectroscopy, atomic force microscope (AFM) characterization, as well as the Time-of-Flight-Secondary-Ion-Mass-Spectroscopy (ToF-SIMS). Transport measurements were performed  to check the superconductivity of exfoliated FeSe flakes.

\section{Nanosheets and Device Fabrication}

A polydimethylsiloxane (PDMS)-based exfoliation method (similar to the scotch tape method developed in the graphene research) was used to produce nanosheets and transfer them onto different substrates\cite{pdms1}. Low-residue Nitto tapes were used to exfoliate FeSe crystals. Then, the Nitto tapes were peeled off from pieces of PDMS films. Some nanosheets were left on the PDMS films and were then transferred to 285 nm SiO$_{2}$/(100) Si substrates. This is a simple and effective way to obtain approximately 10 nm-thick flakes. The typical size of the exfoliated nanosheets is 10 to 20 microns. AFM (Asylum Research model MFP3D) was used to characterize the thicknesses. The colors of fresh nanosheets are yellow for sheets thicker than 60 nm and purple for those thinner than 20 nm. Both optical and AFM images show that the surfaces of the nanosheet are flat and smooth (Fig.1).

\begin{figure}[hbtp]
%\centering
\begin{center}
	\includegraphics[width=1\textwidth]{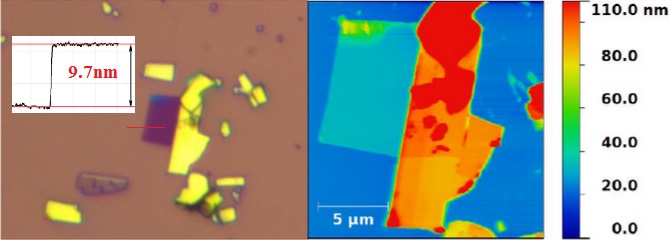}
\end{center}
%\textbf{Figure 1.} 
\textbf{Figure 1.}  Optical and AFM images of FeSe nanosheets. The purple regions are nanosheets less than 20 nm thick, and the yellow regions are FeSe sheets more than 60 nm thick. The insert is an AFM scan showing a thickness of 9.7 nm. The AFM scan location was near the bottom of the dark purple region, as shown by a red line.

%\caption{Optical and AFM images of FeSe nanosheets. In (a), the purple regions are nanosheets less than 20 nm thick, and the yellow regions are FeSe sheets more than 60 nm thick. The insert in (a) is an AFM scan showing a thickness of 9.7 nm. The AFM scan location was near the bottom of the dark purple region, as shown by a red line in (a).}
\label{fig8p5}
\end{figure}

 For the superconductivity transition temperature $T_{c}$ measurements, since the volume of a nanosheet flake is too small for bulk magnetometry measurements, transport measurements were used. For the conductivity characterization, at least two electrodes  need to be added to a nanosheet as the source and drain. A conventional electron beam lithography (EBL) method was first used to pattern these contacts. However, EBL caused serious degradation of the FeSe flakes such that they no longer conducted. The degraded FeSe flakes have the same color as the heavily-oxidized ones. We postulate that some chemical reactions happened during the EBL process, which degraded the FeSe nanosheets. To avoid the degradation and heating, two other methods, transferring onto pre-defined electrodes method and shadow mark evaporation method, were used for the micro-electrodes fabrication. Both of the two methods maintained the conducting properties of FeSe. The two methods are described below.
 %The three methods discussed above are illustrated in Fig.~\ref{fig8p6} and the latter two are described below: 

\begin{enumerate}

\item Method 1: shadow mask evaporation\cite{p4}. 

\begin{enumerate} 

\item Make nanosheets by exfoliation.

\item Use a mask aligner to align a shadow mask on top of a flake. The shadow mask was then fixed with respect to the FeSe/SiO$_{2}$/Si stack. The micron-sized apertures on the shadow mask were made with a Focused Ion Beam Microscope (FIB). 

\item Put the whole setup in a metal evaporator and evaporate gold through the shadow mask apertures to form the contacts.

\end{enumerate}

\item Method 2: transferring thin FeSe flakes onto pre-fabricated electrodes.

\begin{enumerate}

\item Exfoliate FeSe with a Nitto tape and press it onto a piece of PDMS. 

\item Position the flake-baring PDMS piece with a micro-manipulator and transfer flakes onto the pre-fabricated electrodes by gentle touching. 

\end{enumerate}

\end{enumerate}

\section{Air stability}
After obtaining the FeSe nanosheets by exfoliation, we used AFM and micro-Raman spectroscopy to monitor the color change and the change of the phonon modes immediately. ToF-SIMS was used to characterize the chemical composition change.

\subsection{Raman Spectroscopy}

\begin{figure}[!htb]
%\centering
%\subfigure[\label{fig8p9(a)}]
\begin{center}

  {\includegraphics[width=0.65\linewidth,height=5.5cm]{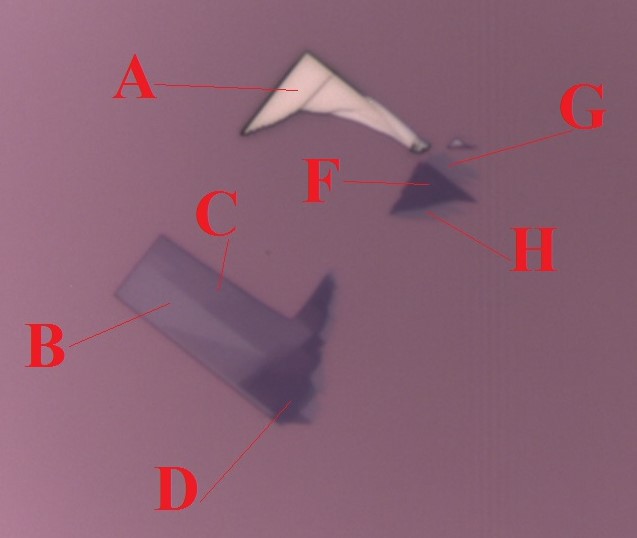}}\hfill
%\subfigure[\label{fig8p9(b)}]

  {\includegraphics[width=0.8\linewidth,height=7cm]
  {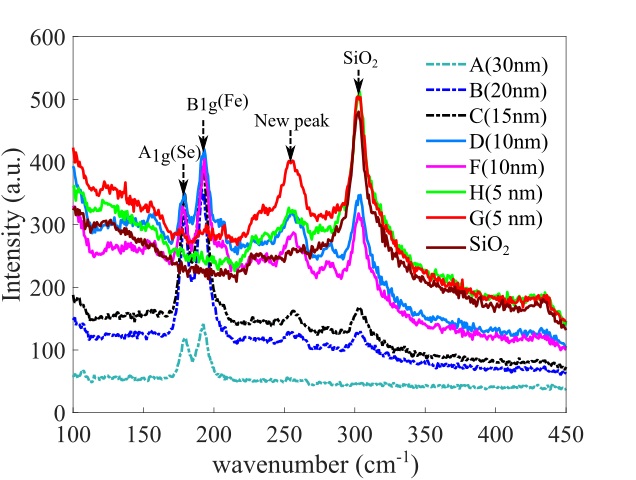}}%\hfill
  
  \end{center}
 % {graphs/170518f.png}}\hfill
%\subfloat[Caption for subfigure 3\label{fig:test3}]

%\caption
%\caption{(a) Optical image of the exfoliated FeSe nanosheets A to G.  (b) Raman spectra of the flakes A to G after 1hr air exposure.}
\textbf{Figure 2.} (a)(top) Optical image of the exfoliated FeSe nanosheets A to G.  (b)(bottom) Raman spectra of the flakes A to G after 1hr air exposure.

\label{fig8p9}
\end{figure}

All Raman measurements were performed with a Horiba Scientific LabRam HR800 Raman system in a backscattering configuration with 632.8 nm excitation wavelength. The Raman spectra were collected through an
Olympus 50X (NA=0.55) objective len and recorded with a grating of 2400 lines/mm which has a spectral resolution of 0.27 cm\textsuperscript{-1}. 
%[[Rui, please confirm the Raman measurement condtions]]
Fig.2a (top) shows the optical image of several flakes with different thicknesses, the thinnest flake is about 5 nm thick, according to the AFM measurements. Fig.2b (bottom) shows the Raman spectra of FeSe flakes A to G after 1 hr air exposure. The two peaks around 200 cm\textsuperscript{-1} are the characteristic peaks of the FeSe lattice\cite{feseraman1}. They are related to the phonon-modes of $P4/nmm$ lattice. The left and the right peaks correspond to the $A_{g}(Se)$ and the $B_{g}(Fe)$ peaks from the vibration of Se and Fe atoms, respectively. We also observed that as the thickness decreases, the flakes become more and more transparent, and a peak around 300 cm\textsuperscript{-1} from the underlying SiO$_2$ appears and strengthens. Besides this peak, a new peak around 250 cm\textsuperscript{-1} also emerges. Meanwhile, the characteristic peaks of the FeSe lattice diminish with decreasing thickness. There are no signs of the characteristic peaks of FeSe lattice in the 5nm-thick flakes. 

 The first question we investigated was the causes of the diminishing of the characteristic FeSe peaks and the emergence of the new peak around 250 cm\textsuperscript{-1}. One possible explanation is the degradation during the air exposure. If this was the case, long time air exposure should introduce this new Raman peak to thick flakes. To check this hypothesis, we performed a 24 hour aging test in open air and found evidence supporting this explanation. Fig.3 shows the change of the Raman spectra of FeSe nanosheets during a prolonged exposure in air. Fig. 3(a)(left) is for a piece of bulk crystal thicker than 300 nm, and Fig. 3(b)(right) is for a 10nm-thick nanosheet. In Fig. 3(a), we can see that the peak at 250 cm\textsuperscript{-1} strengthened greatly in the bulk crystal after a 24 hr exposure in air. The degradation of the FeSe flakes in air can also be observed from the color changes. Nanosheets less than 50 nm thick show a light purple color. After 3 and 24 hr air exposure, the color changed to light green and dark green, respectively.
 
 %, as seen in Fig.~\ref{fig8p11}(a).

\begin{figure}[!htb]
%\centering
%\figure[\label{fig8p10a}]
%width=1.0\textwidth, height=5.5cm
\begin{center}
  {\includegraphics[width=1.0\textwidth]{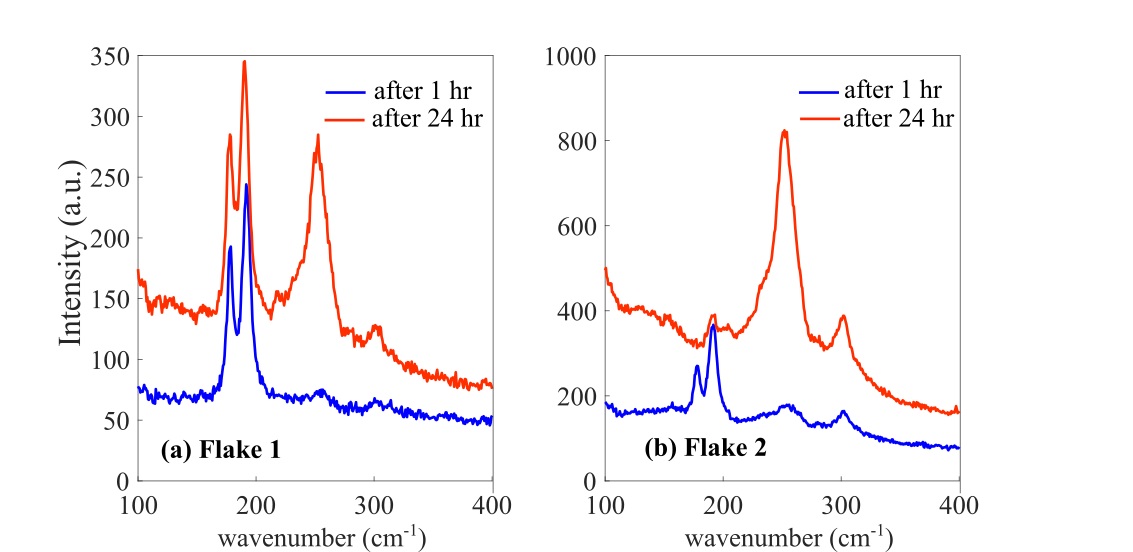}}
  
  %\hfill
  
\end{center}

%\subfigure[\label{fig8p10b}]
%\caption
%\caption{Raman spectra before and after long time exposure in air: (a) a piece of bulk crystal; (b) a nanosheet about 10 nm thick.}
\textbf{Figure 3. }Raman spectra before and after long time exposure in air: (a) a piece of bulk crystal; (b) a nanosheet about 10 nm thick.

\label{fig8p10}
\end{figure}

\subsection{Morphology by AFM}

The next question investigated was the microscopic origin of the peak around 250 cm\textsuperscript{-1}. By comparing with Raman peaks of relevant materials, we found that this peak position is consistent with the Raman peak of the amorphous Se nanoparticles\cite{feseraman2}.
Further evidence of these nanoparticles being Se nanoparticles was found from the AFM characterizations, nanoparticles can be indeed detected on the surface of the nanosheets exposed in air. Fig.4 shows the AFM images taken before and after 24 hr air exposure for FeSe nanosheets. The left AFM image of Fig. 4 shows the surface of a fresh FeSe nanosheet, which is relatively clean and flat. The right AFM image of Fig. 4 shows the same surface after 24 hr air exposure, where particles of several nanometers in diameter emerge on the surface. These particles are considered to be the Se nanoparticles revealed from the Raman data.

\begin{figure}[!htb]
%\centering
%\subfigure[\label{fig9p10b}]
\begin{center}
%{\includegraphics[width=0.5\linewidth]
%{Fig5a.jpg}}

%\subfigure[\label{fig9p10b}]
{\includegraphics[width=1\linewidth]
{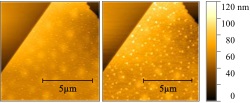}}

%\hfill
\end{center}

%\caption
%\centering
%\caption{(a) Optical images of 3 FeSe flakes before (top row) and after (bottom row) 24 hr air exposure, respectively. (b) AFM images of a typical FeSe flake before (left) and after (right) 24 hr air exposure, respectively. }
\textbf{Figure 4.} AFM images of a typical FeSe flake before (left) and after (right) 24 hr air exposure, respectively. 
\label{fig8p11}
\end{figure}
\subsection{ToF-SIMS}
The third question we addressed was about the chemical processes involved during the air exposure. Raman or optical images were incapable to tell the details of the chemical change, especially during the first hour of air exposure. 
%The most obvious conjecture about this chemical change would be the oxidation of FeSe.

To address this, we performed a precise chemical composition analysis of the FeSe nanosheets. The ToF-SIMS instrument used was PHI Trift V nanoToF-SIMS model\cite{2d1}. The nanoimaging and depth profiling can give the lateral and depth distribution of the chemical compositions. In our ToF-SIMS measurements (see Fig.5), we characterized the chemical compositions for FeSe nanosheets
before and after 24 hr exposure in air. For nanosheets exposed in air for less than 20 mins, we found that the nanosheets remain intact, and no obvious oxidation was found. For nanosheets exposed in
air for about 2 hours, no significant signal of oxides were detected. For nanosheets exposed in air for 24 hr, iron oxides were found in the top 5 nm of the nanosheets. The composition of the nanosheets exposed in air is also determined, several iron oxides (such as FeO) were detected. This suggests that oxidation process occurs during air exposure, which is considered to be the reason for the precipitation of Se nanoparticles.

\begin{figure}[ptb]

\centering
\begin{center}
\includegraphics[scale=1]{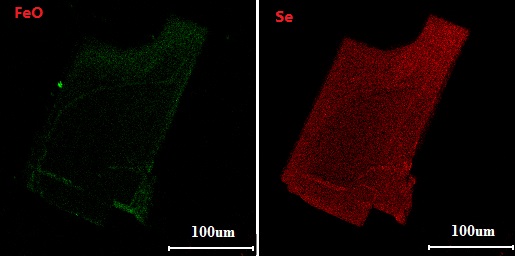}
%\caption
\end{center}
%\hfill
%\centering
%\caption{ToF-SIMS images for a FeSe flake exposed in air for 24 hr showing the FeO and Se contents.}
\textbf{Figure 5. }ToF-SIMS images for a FeSe flake exposed in air for 24 hr showing the FeO and Se contents.
\label{fig8p12}
\end{figure}

\subsection{Superconductivity measurements}

The last question investigated was the superconductivity and the gate modulation of the exfoliated nanosheets. As the flakes are too small for the bulk magnetometry, a transport method was used to measure the transition temperature $T_c$. The electrodes fabrication details were described in Section 2. High resistance was measured in FeSe flakes with more than 24 hr air exposure. However, for FeSe nanosheets with less than 2 hr air exposure, the superconductivity was retained. We performed transport measurements on 6 flakes with the thicknesses ranging from 10 to 60 nm. 4 of these flakes were found to be superconducting. However, compared to the bulk FeSe sample of 300 nm thickness, the superconducting transition of these exfoliated FeSe nanosheets was not as sharp and $T_{c}$ was reduced to 5 to 7 K, as seen in Table I and Fig.6. The lower $T_{c}$ and broadened superconducting transition are consistent with the fact that these FeSe nanosheets were oxidized. Gating by SiO$_2$ was also attempted on the 10nm-thick flake. No gating effect was found, thus the carrier density of the FeSe nanosheet is way beyond the manageable range that a SiO$_2$ gate can introduce.

\begin{figure}[!htb]
%\centering
\begin{center}
%\captionsetup{width=\textwidth}
\includegraphics[width=1.0\textwidth]{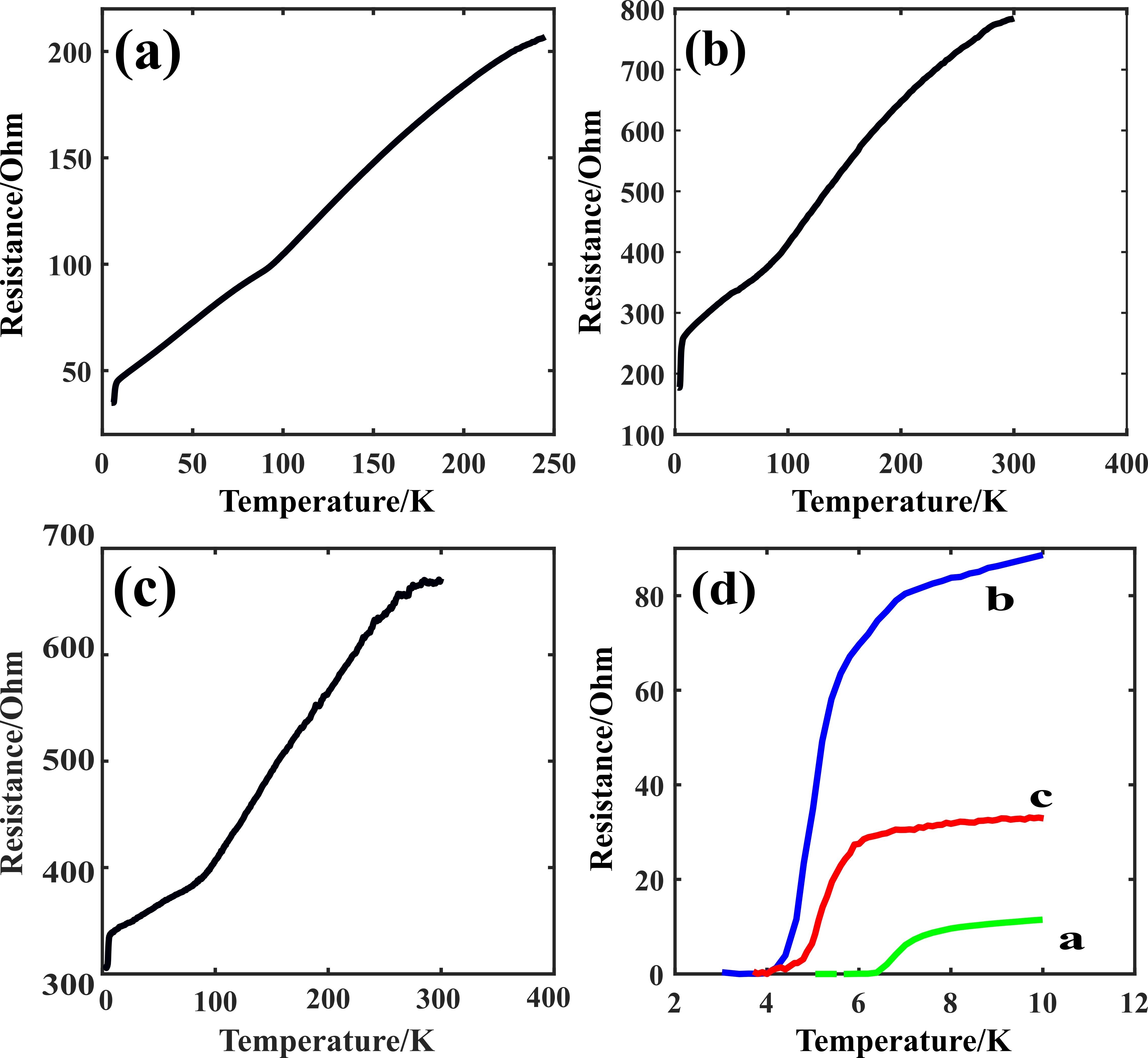}
%1by3e2.png
%\begin{minipage}[t]{17cm}

\end{center}

%\caption{Resistance as a function of temperature of (a) a 60 nm thick flake; (b): a 15 nm thick flake; and (c): a 10 nm thick flake. (d): The comparison of the samples in a to c for T$<$10 K. The contact resistance was subtracted.}
\textbf{Figure 6.}
Resistance as a function of temperature of (a) a 60 nm thick flake; (b): a 15 nm thick flake; and (c): a 10 nm thick flake. (d): The comparison of the samples in a to c for T$<$10 K. The contact resistance was subtracted.
\label{fig8p13}
%\end{minipage}
\end{figure}

\begin{table}[ht]
%\label{table8p1}
\centering
%\caption{$T_{c}$ and thicknesses measured on 7 samples.}
\textbf{Table I: }$T_{c}$ and thicknesses measured on 7 samples.

%\begin{tabular}{c{4cm}|g{4cm}|g{4cm}}

\begin{tabular}{|p{2.0cm}|p{3cm}|p{2.0cm}|  }

\hline

&$T_{c}$  &Thickness \\
\hline

bulk   & \ 9 K & \ -- \\
flake 1& \ 7 K & \  $\mathtt{\sim}$ 60 nm \\
flake 2& \ 6 K & \  $\mathtt{\sim}$ 15 nm \\
flake 3& \ 5 K & \  $\mathtt{\sim}$ 10 nm \\

flake 4& \ 6 K & \  $\mathtt{\sim}$ 15 nm \\

flake 5& \ semiconducting & \  $\mathtt{\sim}$ 10 nm \\

flake 6& \ semiconducting & \  $\mathtt{\sim}$ 10 nm \\

\hline
\end{tabular}

\label{table8p1}

\end{table}

\section{Conclusion}

% 	In summary, this work reported on the layer-by-layer sublimation of black phosphorus at 500 - 600 K. The thinning rates for "continuous heating" were $\sim$ 0.18 nm/min and 1.15 nm/min at 500 and 550 K, respectively. The Raman intensity ratios of $\frac{I_{Si}}{I_{A_g^2}}$ as functions of BP thickness at room temperature, 500 K and 550 K were measured, which could be used as an in-situ and non-contact determination of BP thickness and thickness control during the sublimation thinning process. Large (with areas $>$ 200 $\mu$m$^2$ ) and few-layer (2 to 4 layers) BP flakes with good integrity, uniformity and crystallinity on Si wafer and graphene/Si substrate were prepared successfully and repeatedly. No micron scale defects were observed.  We expect this method also works for deposited BP besides exfoliated BP. The sublimation thinning method is promising in further fabrication of high quality few-layer BP in large scale.
In summary, this work investigated the degradation of exfoliated FeSe nanosheets in air. Raman peaks of Se particles emerged after air exposure, and the oxidation of iron was also observed. Further, FeSe also degrades in the conventional EBL process for making electrodes, and two alternative methods have been developed for making electrodes. Transport measurements showed a reduced $T_{c}$ and a broadened transition in the nanosheets. Our study provides a clearer picture of the stability of FeSe nanosheets when they are exposed in air. The knowledge obtained paves the way for building FeSe nanodevices with higher quality. It could also be useful for a thorough understanding of the experimental results based on the FeSe nanosheets exposed in air, such as the ion-gating experiment\cite{fese1}.
%\section*{Author contributions}

% W. L. and G. X. conceived the experiments. W. L conducted more than 90\% of the thermal thinning, Raman measurements and analyzed the results. W. L. prepared the majority of the samples and performed the majority of the AFM measurements. R. Y. and Y. Zhao contributed by some AFM and Raman measurements respectively. J. Liu and W. Zhu contributed by providing initial samples and helpful discussions. W. L. and G. X. led the writing of the paper, and all the authors participated in the discussions of results. The project was supervised by G. X. All authors reviewed the manuscript. 

\section*{Acknowledgement}
The authors thank John Kim and Philip Wong from the Interfacial Analysis and Reactivity Laboratory, the University of British Columbia (UBC) for helpful discussions. Professor Josh Folk from the Department of Physics and Astronomy, UBC is acknowledged for providing the lab apparatus and Teren Liu from the Department of Materials Engineering, UBC for his assistance in preparing the figure images. This work has been supported by The Natural Sciences and Engineering Research Council of Canada (NSERC),  the Canada Foundation for Innovation (CFI), and the Stewart Blusson Quantum Matter Institute (SBQMI) at UBC.
% The authors acknowledge Dr. David Tuschel (Horiba Scientific, Edison, NJ, U.S.A) for discussions on angle-resolved polarized Raman spectroscopy. The authors acknowledge Chris Balicki (4D LABS, Simon Fraser University, Burnaby, BC, Canada) for his assistance in Atomic Force Microscopy and Ms. Zenan Jiang (Electrical Engineering, University of British Columbia, Vancouver, BC, Canada) for supplying the CVD graphene.

%\section*{Additional information}
%\textbf{competing financial interests: }All other authors declare no competing financial
%interests
%\par

\section*{Reference}
\bibliography{mybib.bib}

\newpage

\vspace{10pt}

\end{document}